\documentclass[prb,aps,superscriptaddress,twocolumn,amsmath,amssymb,showpacs]{revtex4-2}
\usepackage{amssymb}
\usepackage{booktabs}
\usepackage{multirow}% ±í¸ñºÏ²¢
\usepackage{epstopdf}
\usepackage{float}
\usepackage{graphicx}% Include figure files
\usepackage{dcolumn}% Align table columns on decimal point
\usepackage{bm}% bold math
\usepackage{color}
\usepackage{ulem}

\hfuzz=\maxdimen
\begin{document}	

  \title{Inter-Cation Charge Transfer Mediated Antiferromagnetism in Co$_{1+x}$Ir$_{2-x}$S$_4$}
	
	\author{Liang-Wen Ji} \email[]{lwji@zju.edu.cn}
	\affiliation{School of Physics, Zhejiang University, Hangzhou 310058, China}

	\author{Si-Qi Wu}
	\affiliation{School of Physics, Zhejiang University, Hangzhou 310058, China}

	\author{Bai-Zhuo Li}
	\affiliation{School of Physics, Zhejiang University, Hangzhou 310058, China}
	
	\author{Wu-Zhang Yang}
	\affiliation{School of Sciences, Westlake Institute for Advanced Study, Westlake University, Hangzhou 310064, China}
	
	\author{Shi-Jie Song}
	\affiliation{School of Physics, Zhejiang University, Hangzhou 310058, China}
	
	\author{Yi Liu}
	\affiliation{School of Physics, Zhejiang University, Hangzhou 310058, China}
	\affiliation{Department of Applied Physics, Zhejiang University of Technology, Hangzhou 310023, China}
	
	\author{Jing Li}
	\affiliation{School of Physics, Zhejiang University, Hangzhou 310058, China}
	
	\author{Zhi Ren}
	\affiliation{School of Sciences, Westlake Institute for Advanced Study, Westlake University, Hangzhou 310064, China}
	
	\author{Guang-Han Cao} \email[]{ghcao@zju.edu.cn}
    \affiliation{School of Physics, Zhejiang University, Hangzhou 310058, China}
    \affiliation{Interdisciplinary Center for Quantum Information, and State Key Laboratory of Silicon and Advanced Semiconductor Materials, Zhejiang University, Hangzhou 310058, China}
    \affiliation{Collaborative Innovation Centre of Advanced Microstructures, Nanjing University, Nanjing, 210093, China}
	
	\date{\today}% It is always \today, today,
	%  but any date may be explicitly specified
	
\begin{abstract}
The antiferromagnetism in transition metal compounds is mostly mediated by the bridging anions through a so-called superexchange mechanism. However, in materials like normal spinels $AB_2X_4$ with local moments only at the $A$ site, such an anion-mediated superexchange needs to be modified. Here we report a new spinel compound Co$_{1+x}$Ir$_{2-x}$S$_4$ ($x$ = 0.3). The physical property measurements strongly suggest an antiferromagnetic-like transition at 292 K in the Co($A$) diamond sublattice. The first-principle calculations reveal that the nearest-neighbor Co($A$) spins align antiferromagnetically with an ordered magnetic moment of 1.67 $\mu_\mathrm{B}$, smaller than the expected $S = 3/2$ for Co$^{2+}$. In the antiferromagnetic state, there exists an inter-cation charge-transfer gap between the non-bonding Ir-$t_\mathrm{2g}$ orbitals at the valence band maximum and the Co-S antibonding molecular orbitals at the conduction band minimum. The small charge transfer energy significantly enhances the virtual hopping between these two states, facilitating a robust long-range superexchange interaction between two neighboring CoS$_4$ complexes, which accounts for the high N\'{e}el temperature in Co$_{1+x}$Ir$_{2-x}$S$_4$. This inter-cation charge transfer mediated magnetic interaction expands the traditional superexchange theory, which could be applicable in complex magnetic materials with multiple cations.
\end{abstract}

%74.70.Xa Pnictides and chalcogenides
%75.30.-m Intrinsic properties of magnetically ordered materials
%61.66.Fn Inorganic compounds
%72.80.Ga Transition-metal compounds
%71.45.-d Collective effects
%75.50.Ee Antiferromagnetics
%71.20.Nr Semiconductor compounds
\maketitle
\section{\label{sec:level1}Introduction}

In transition metal compounds, strong electron-electron interactions often give rise to localization of electrons, resulting in Mott insulators~\cite{1-imada-1998}. According to the Zaanen-Sawatzky-Allen scheme, Mott insulators can be classified into Mott-Hubbard type ($U < \Delta$) and charge-transfer type ($U > \Delta$), depending on the on-site Coulomb energy $U$, the charge-transfer energy $\Delta$, and the transfer integrals $t$~\cite{2-Zaanen-1985}. In Mott insulator, particularly within the charge-transfer type, the localized spins are antiferromagnetically coupled through a superexchange (SE) interaction~\cite{3-anderson-1959}. In the SE, the nonmagnetic (NM) ligand mediates the exchange interaction between the two spins on the cations~\cite{4-anderson-1950}. With the wide applicability of the Goodenough-Kanamori-Anderson (GKA) rule~\cite{5-goodenough-1963}, SE has been recognized as the dominant mechanism governing indirect magnetic exchange through intervening nonmetal anions in Mott insulators. While in polyvalent anion systems, SE interactions mediated by different anions are modified, termed extended SE interactions, exhibiting inconsistent results with the GKA rules~\cite{6-zhang-2019,7-Jiang-2021}. Additionally, a super-SE interaction mediated by electron hopping through two anions has been proposed to explain the exchange interactions between localized spins in van der Waals materials~\cite{8-sivadas-2018,9-huang-2020}. However, the applicability of SE in complex scenarios, categorized as long-range SE, where magnetic ions are separated over a long path comprising multiple anions and even cations, remains insufficiently explored~\cite{10-roth-1964,11-geertsma-1990}. 

Spinel system provides an ideal platform for investigating the long-range SE effects~\cite{10-roth-1964,12-roth-1964}. With the formula $AB_2X_4$, the spinel structure adopts a close-packed face-centered cubic arrangement of anions ($X$), facilitating two distinctive cationic positions: the tetrahedrally coordinated $A$-site and the octahedrally coordinated $B$-site~\cite{13-tsurkan-2021}. Magnetic ions can occupy either site, leading to complex magnetic exchange interactions within spinels~\cite{14-Lago-2010,15-guratinder-2019,16-guratinder-2022,17-yaresko-2008}. Specifically, when magnetic ions are located on the $A$-site, the $A$-$A$ magnetic exchange pathway indirectly involves at least three intermediate ions~\cite{12-roth-1964,13-tsurkan-2021}. Generally, inherent frustration of $A$-site magnetic spinel limits magnetic ordering to low temperatures~\cite{18-bergman-2007}. For instance, strong frustration leads to the emergence of spiral spin-liquid behavior below 2 K in MnSc$_2$S$_4$~\cite{19-gao-2017} and the suppression of magnetic ordering temperature in FeSc$_2$S$_4$~\cite{20-fritsch-2004,21-Chen-2009,22-plumb-2016,23-tsurkan-2017}. However, despite weak frustration, CoAl$_2$O$_4$ exhibits AFM order around 10 K~\cite{24-roy-2013}, likely due to the weak long-range exchange interaction~\cite{10-roth-1964,13-tsurkan-2021}. 

In contrast to CoAl$_2$O$_4$~\cite{24-roy-2013}, Co$_3$O$_4$ and CoRh$_2$O$_4$ are typical antiferromagnets with elevated N\'{e}el temperatures ($T_\mathrm{N}$) of 40 K and 25 K, respectively~\cite{25-scheerlinck-1976,26-GeL.-2017}. In particular, CoRh$_2$S$_4$ is an antiferromagnetic (AFM) insulator with a $T_\mathrm{N}$ of 418 K, distinct from other cobalt thiospinels~\cite{27-Ohashi-1987}. The varying N\'{e}el temperature with different $B$-site cations suggests the participation of $B$-site ions in the long-range exchange interaction~\cite{12-roth-1964,28-mayer-1981}. So far, few studies have delved into the underlying variations in the exchange interactions. Mayer et al.~\cite{28-mayer-1981} gave an incomplete description of the evolution of exchange interactions. Geertsma et al.~\cite{11-geertsma-1990} proposed a path-model using two-particle Green functions, yet without considering the NM $B$-site cations. Thus, a comprehensive understanding of the mechanism of long-range SE is yet to be achieved. 

%XRD 
\begin{figure}[t]
	\includegraphics[width=8.5cm]{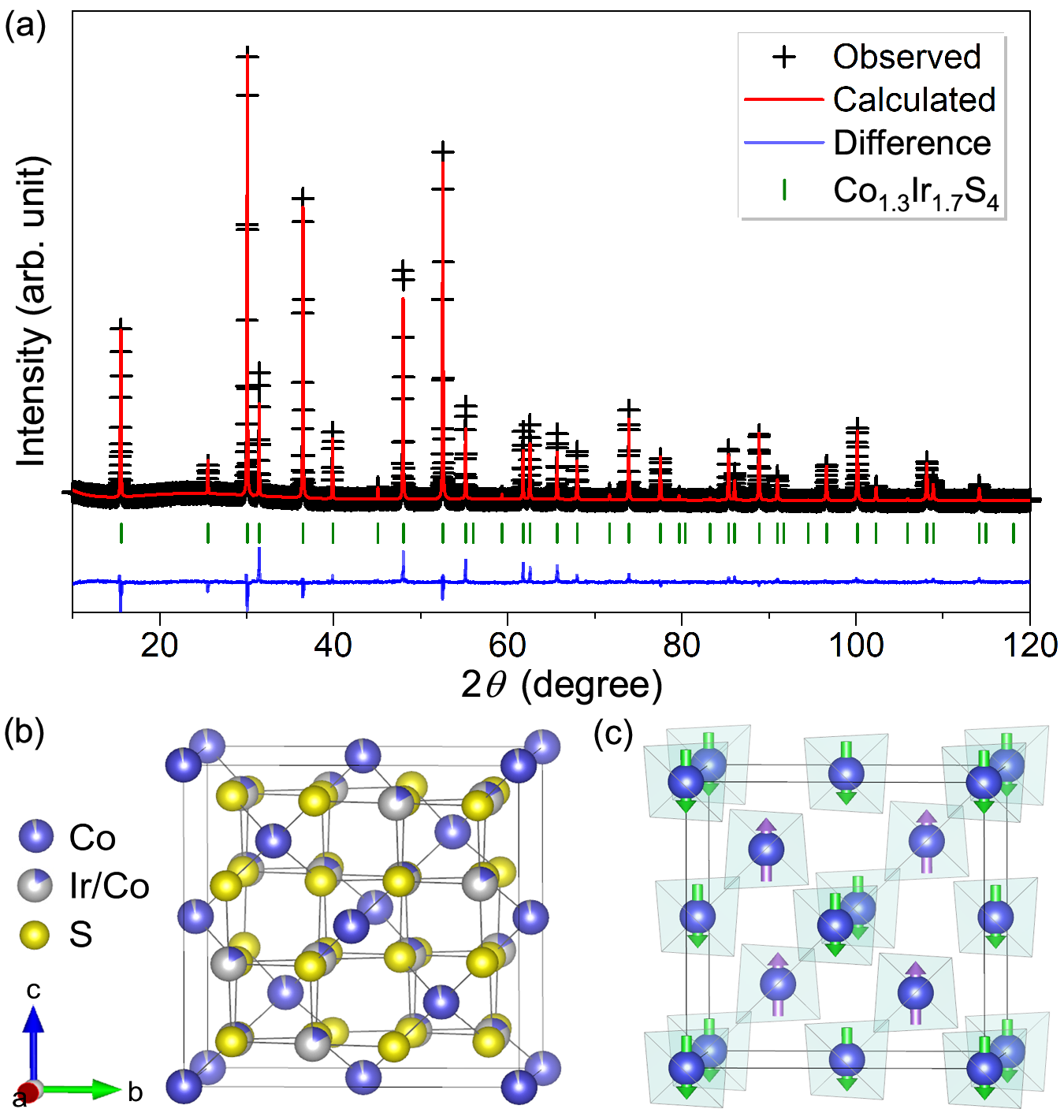}
	\caption{(Color online) (a) Powder X-ray diffraction with the Rietveld refinement profile for Co$_{1+x}$Ir$_{2-x}$S$_4$ ($x$ = 0.3) polycrystalline. (b) The refined crystal structure of Co$_{1.3}$Ir$_{1.7}$S$_4$. (c) The diamond sublattice antiferromagnetic structure with purple and green arrows representing up and down spins, respectively.} 
	\label{XRD}
\end{figure}

To clarify the long-range SE mechanism within Co($A$) spinels, where cobalt occupies the $A$ site, we synthesized a new spinel Co$_{1+x}$Ir$_{2-x}$S$_4$. An ionic configuration of Co$^{2+}$[(Co$^{3+}$)$_x$(Ir$^{3+}$)$_{2-x}$](S$^{2-}$)$_4$ is assigned, with Co$^{2+}$ in a high-spin state ($S$ = 3/2) at the $A$ site and non-magnetic Co$^{3+}$ and Ir$^{3+}$ at the $B$ site, according to the crystal field theory~\cite{29-burns-1993}. The electrical, magnetic, and thermodynamic properties were measured, which indicate that Co$_{1+x}$Ir$_{2-x}$S$_4$ is an AFM insulator with $T_\mathrm{N}$ $\sim292$ K. Theoretical calculations were also utilized to explore the magnetic and electronic properties. Strong hybridization, as confirmed by crystal orbital Hamilton populations (COHP) and partial charge density, results in a reduced local moment of Co($A$) and a more precise representation by molecular orbitals (MOs) in this system. Careful analysis of the electronic structures reveals an unusual inter-cation charge-transfer (ICCT) energy gap between the antibonding states of Co-$t_2$ and S-$p$ (${t_2}^\mathrm{Co}$-$p^\mathrm{S\ast}$) MOs and Ir-$t_\mathrm{2g}$ orbitals. The narrow charge-transfer gap fosters significant electron hopping between the ${t_2}^\mathrm{Co}$-$p^\mathrm{S\ast}$ and Ir-$t_\mathrm{2g}$ states. Consequently, the long-range SE interaction is established, wherein Ir-$t_\mathrm{2g}$ orbitals mediate the AFM coupling between two nearest CoS$_4$ complexes. Moreover, the validity of long-range SE is further confirmed in other cobalt spinel systems.

\begin{table}[b]
	\caption{Crystallographic data for Co$_{1.3}$Ir$_{1.7}$S$_4$ at room temperature from the Rietveld refinement of powder X-ray diffraction.}
	\renewcommand\arraystretch{1.2}
	\begin{tabular}{cccccc}
		\hline\hline % Top horizontal line
		\multicolumn{3}{c}{Chemical formula} & \multicolumn{3}{c}{Co$_{1.3}$Ir$_{1.7}$S$_4$} \\
		\multicolumn{3}{c}{Crystal system} & \multicolumn{3}{c}{Cubic} \\
		\multicolumn{3}{c}{Space group} & \multicolumn{3}{c}{$F$$d\bar{3}m$ (origin choice 1)} \\
		\multicolumn{3}{c}{$a$ ({\AA})} & \multicolumn{3}{c}{9.8399(9)} \\
		\multicolumn{3}{c}{$V$ ({\AA}$^{3}$)} & \multicolumn{3}{c}{952.7(3)} \\
		\multicolumn{3}{c}{$R_{\rm wp}$ (\%)} & \multicolumn{3}{c}{5.56} \\
		\multicolumn{3}{c}{$R_{\rm p}$ (\%)} & \multicolumn{3}{c}{3.52} \\
		\multicolumn{3}{c}{$S$} & \multicolumn{3}{c}{1.93} \\
		\hline % In-table horizontal line
		Atoms & site & $x$ & $y$ & $z$ & Occupancy  \\
		Co(1) & 8$a$ & 0.2500 & 0.2500 & 0.2500 & 0.979(2)    \\
		Ir(1) & 8$a$ & 0.2500 & 0.2500 & 0.2500 & 0.021    \\
		Co(2) & 16$d$ & 0.6250 & 0.6250 & 0.6250 & 0.161   \\
		Ir(2) & 16$d$ & 0.6250 & 0.6250 & 0.6250 & 0.839   \\
		S     & 32$e$ & 0.3821(2) & 0.3821(2) & 0.3821(2) & 1.000   \\
		\hline\hline % Bottom horizontal line	
	\end{tabular}
	\label{Table1}
\end{table}

% Res
\begin{figure*}[t]
	\includegraphics[width=16cm]{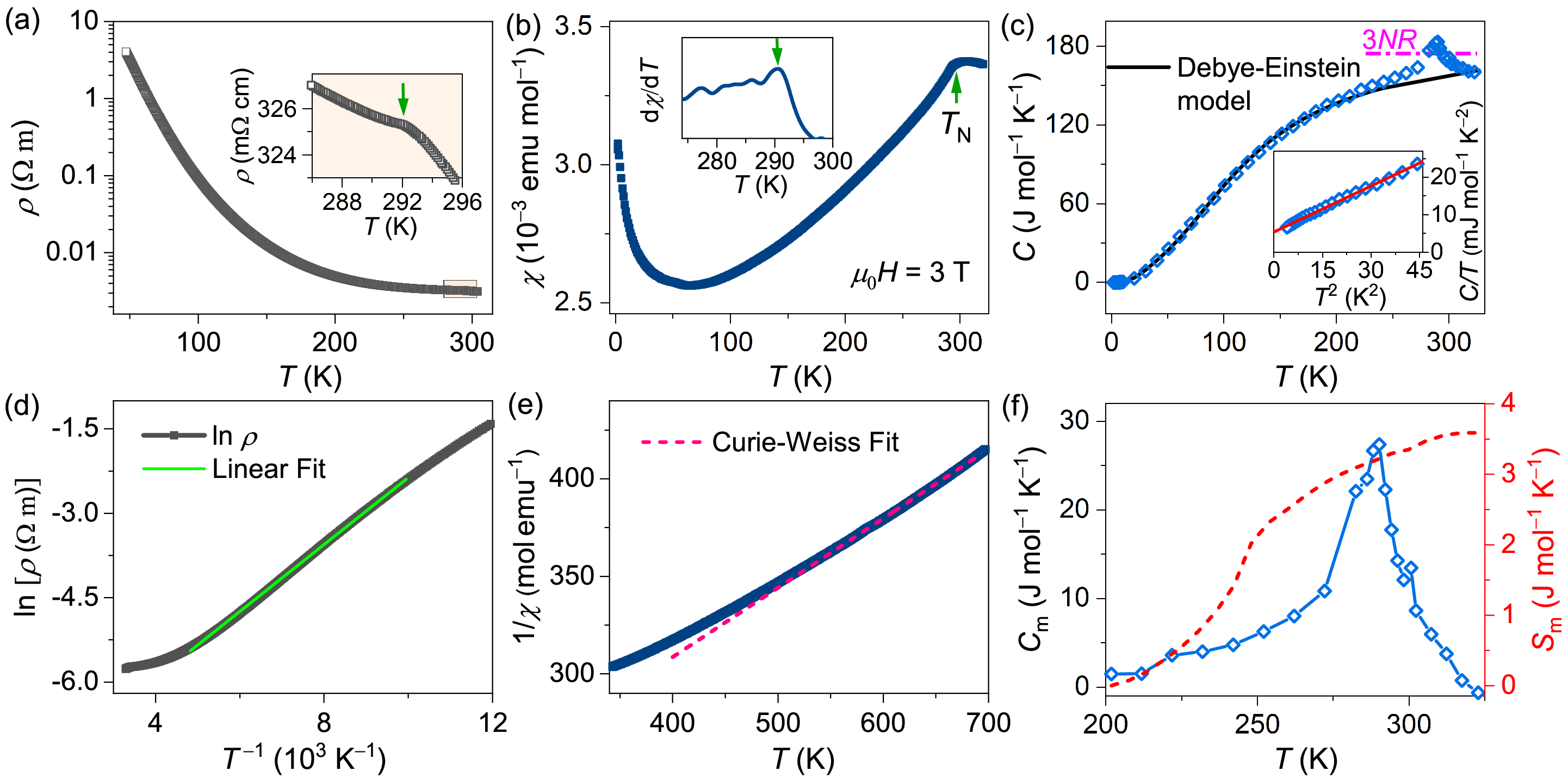}
	\caption{(Color online) Physical properties of the Co$_{1.3}$Ir$_{1.7}$S$_4$ polycrystalline sample. (a) Temperature dependence of electrical resistivity. The inset is a close-up of the resistivity around 290 K. (b) Temperature dependence of magnetic susceptibility under a magnetic field of 3 T. The inset is the first derivative of magnetic susceptibility. (c) Specific heat $C$ as a function of temperature. The inset is the $C/T$ versus $T^2$ plot at low temperatures. (d) The Arrhenius fit of the resistivity. (e) Temperature dependence of the reciprocal of high-temperature magnetic susceptibility. The pink line represents a fit with the Curie-Weiss formula. (f) The magnetic contribution (estimated by subtracting the lattice contribution) $C_\mathrm{m}$ (left axis) and the magnetic entropy $S_\mathrm{m} =\int_{0}^{T}\frac{C_\mathrm{m}}{T}dT$ (right axis) as functions of temperature.} 
	\label{Exp}
\end{figure*}

\section{\label{sec:level2}Methods}

The Co$_{1+x}$Ir$_{2-x}$S$_4$ polycrystalline sample was synthesized by solid-state reactions. High purity Co powder (99.9\%), Ir powder (99.95\%), and S powder (99.999\%) starting materials with a mole ratio of ($1+x):(2-x):4$ were well mixed in an agate mortar. The desired mixtures were calcined in sealed evacuated quartz tubes in a box furnace at 1250 K for 48 h. Then, the reacted powder was ground for homogenization in an agate mortar, pressed into pellets, and sintered at the same temperature for another 48 h. Note that an argon-filled glove box was employed for the operations above to avoid the possible reactions with water and oxygen. We found that the sample purity depends on $x$. The sample of $x = 0$ consistently contains significant amount of Ir$_2$S$_3$ impurity, suggesting that Co occupies the $B$ site. At higher cobalt concentrations of $x \geq 0.35$, a secondary phase of CoS$_2$ emerges (see Fig. S1 in the Supplemental Material (SM)~\cite{30-SM}). The phase-pure sample could be obtained for $x$ = 0.3. 

Powder X-ray diffraction (XRD) was carried out at room temperature on a PANalytical X-ray diffractometer with a monochromatic Cu $K_{\alpha1}$ radiation. The crystal structure was refined by a Rietveld analysis using RIETAN-FP~\cite{31-izumi-2007}. We employed a Quantum Design magnetic property measurement system (MPMS-5) to measure the magnetization as a function of temperature and magnetic field. The transport properties were measured by a standard four-terminal method on a vibrating sample magnetometer (VSM). A Quantum Design physical property measurement system (PPMS-9) was employed to measure the specific heat. The resistivity measurement employed a standard four-probe method, and the heat-capacity measurement utilized a thermal relaxation technique. The absorption spectrum was measured using a Fourier-transform infrared spectrometer (Bruker Vertex 70) at room temperature. All the samples measured are from the same sintered pellet of Co$_{1+x}$Ir$_{2-x}$S$_4$ ($x =$ 0.3). 

Our DFT calculations were performed using the projector augmented-wave method~\cite{32-PAW}, as implemented in the Vienna Ab initio Simulation Package (VASP)~\cite{33-VASP}. The exchange-correlation functional constructed by the GGA-type Perdew-Burke-Ernzerhof (PBE) was used~\cite{34-GGA}. To better describe the on-site Coulomb interaction, a Hubbard $U$ correction was applied for the Co 3$d$ states (GGA+$U$)~\cite{35-+U}. The wave functions were expanded in the plane waves basis with an energy cutoff of 450 eV, and a $8 \times 8 \times 8$ and $6 \times 6 \times 6$ $\Gamma$-centered k mesh was adopted for the primitive cell and conventional cell (Co$_{8}$Ir$_{16}$S$_{32}$ and Co$_{10}$Ir$_{14}$S$_{32}$) calculations, respectively. The COHP calculations were performed using the version 4.0.0 of the LOBSTER software~\cite{36-nelson-2020}. All the crystal structure and ionic sites of the primitive cell were fully optimized, while only ionic sites of the supercell were relaxed. The calculated lattice parameter is $a_0=$ 6.8580 \AA, corresponding to the conventional unit cell with $a=\sqrt{2}a_0=$ 9.7002 \AA.

\section{\label{sec:level3}Results and Discussion}

\subsection{\label{subsec:level1}Structure.}
We obtained pure-phase samples of Co$_{1+x}$Ir$_{2-x}$S$_4$ by introducing 30\% excess Co, avoiding the appearance of the Ir$_2$S$_3$ secondary phase. Fig.~\ref{XRD}(a) shows the XRD pattern and the Rietveld refinement of the Co$_{1+x}$Ir$_{2-x}$S$_4$ ($x$ = 0.3) powdered sample. Owing to the common site inversion in most spinels~\cite{37-skvortsova-2002}, we allowed mixed occupancy of Co and Ir between the $A$ and $B$ sites while constraining their nominal composition during refinement. Details of the structural refinement together with the crystal parameters are given in Table~\ref{Table1}. Co$_{1.3}$Ir$_{1.7}$S$_4$ crystallized in a spinel structure (Fig.~\ref{XRD}(b)), with the refined lattice parameter $a$ = 9.8399(9) \AA. The $A$ site is occupied by 97.9\% Co and 2.1\% Ir, indicating a minor degree of site inversion ($\sim$ 2.1\%) for Co$_{1.3}$Ir$_{1.7}$S$_4$. Simultaneously, Ir and the excess Co occupy the $B$ site. Besides, the structural parameter $u$ = 0.3821(2) indicates slight distortion of the Co/IrS$_6$ octahedra along the $\left\langle111\right\rangle$ direction. 

\subsection{\label{subsec:level2}Physical Properties Measurement.}

% Res
The temperature dependence of electrical resistivity $\rho(T)$ for Co$_{1.3}$Ir$_{1.7}$S$_4$ is depicted in Fig.~\ref{Exp}(a). Co$_{1.3}$Ir$_{1.7}$S$_4$ shows a semiconducting behavior and owns a room-temperature resistivity $\rho$(300 K) of 0.0032 $\Omega$ m. From the inset of Fig.~\ref{Exp}(a), one observes a kink-like anomaly at 292 K. The mixed-valence state of Ir in CuIr$_2$S$_4$ leads to spin dimerization, charge ordering, and a concomitant metal-to-insulator transition~\cite{38-radaelli-2002}, while no evidence has been found in our system. The $\rho(T)$ data can be fitted with the Arrhenius model $\ln\rho = \ln\rho_\mathrm{0}$ + $E_\mathrm{a}/k_\mathrm{B}T$, where $k_\mathrm{B}$ is Boltzmann constant (Fig.~\ref{Exp}(d)). The thermal activation energy $E_\mathrm{a}$ obtained is $\sim$0.048 eV, which is slightly larger that of CoRh$_2$S$_4$~\cite{39-Kondo-1976}. To more precisely determine the energy gap, the optical absorption measurement was performed (Fig. S2 in the SM~\cite{30-SM}). Utilizing the Tauc plot method~\cite{40-makula-2018}, we estimated an indirect band gap of 0.053(1) eV, basically consistent with the transport measurement result. 

% Phase-U
\begin{figure}[b]
	\includegraphics[width=8cm]{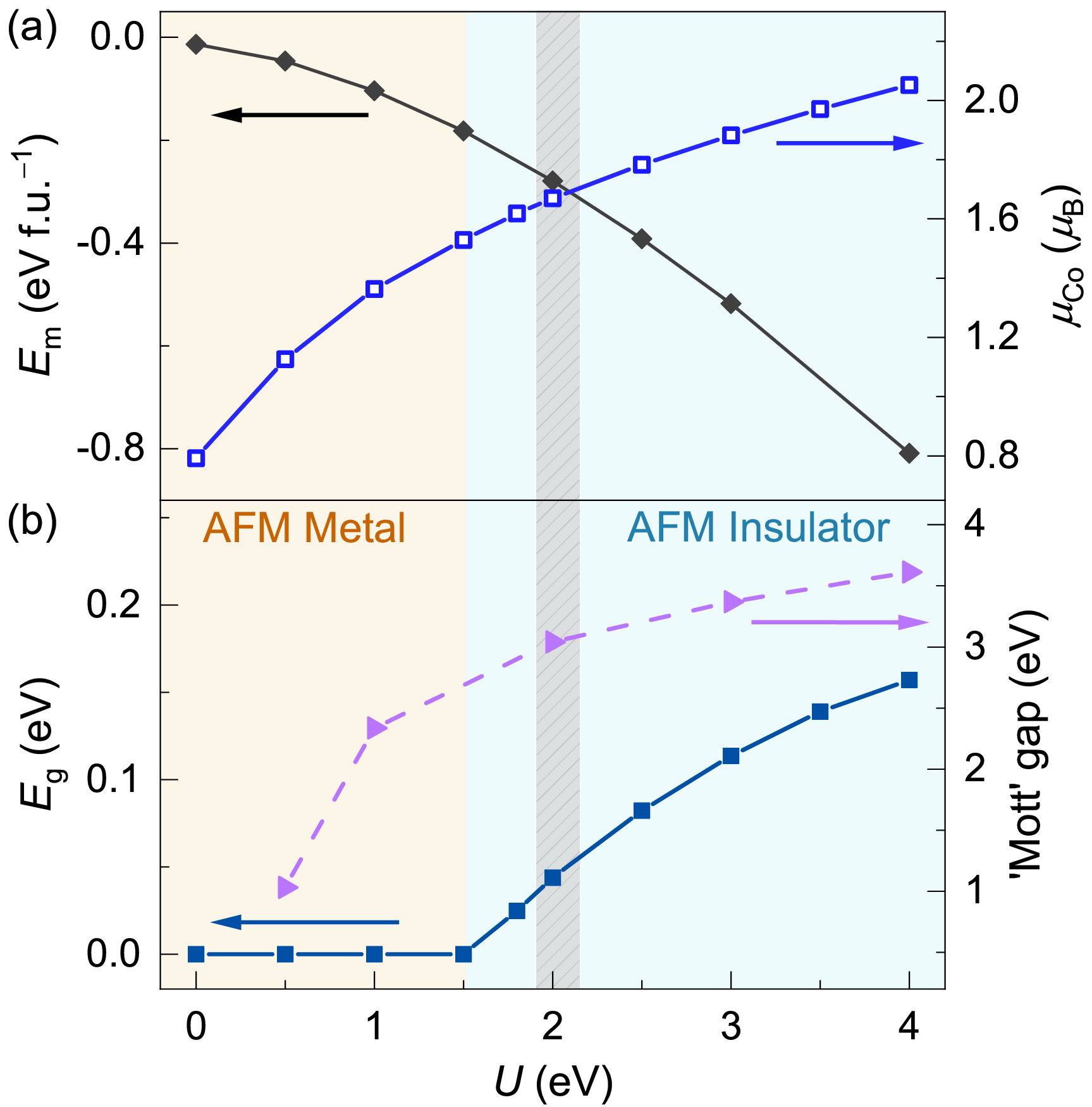}
	\caption{(Color online) (a) Magnetic energy, $E_\mathrm{m}$ = $E_\mathrm{AFM} - E_\mathrm{NM}$, of the antiferromagnetic (AFM) state (left axis) and magnetic moment at the Co($A$) site $\mu_\mathrm{Co}$ (right axis) of CoIr$_2$S$_4$ as functions of $U$ in the GGA + $U$ calculations. (b) Band gap $E_\mathrm{g}$ (left axis) and `Mott' gap (right axis) of CoIr$_2$S$_4$ as functions of $U$. The `Mott' gap is defined in the text. The two ground states, AFM metal and AFM insulator, are distinguished by light orange and light cyan colors respectively. The gray area represents states that are consistent with experimental observations.} 
	\label{Phase_U}
\end{figure}

% Mag
Figure~\ref{Exp}(b) shows the temperature dependence of magnetic susceptibility from 2 K to 320 K for Co$_{1.3}$Ir$_{1.7}$S$_4$. The magnetic susceptibility encounters a steep decrease at around 290 K, suggesting an AFM transition. Below this transition, the magnetic susceptibility decreases with decreasing temperature, and it exhibits a paramagnetic tail at low temperatures. From the derivative of magnetic susceptibility d$\chi$/d$T$, we explicitly determine $T_\mathrm{N} =$ 292 K (inset of Fig.~\ref{Exp}(b)). Note that the sister cobalt thiospinels Co$_3$S$_4$ and CoRh$_2$S$_4$ undergo AFM transitions at 62 K and 418 K~\cite{41-ji-2022,27-Ohashi-1987}, respectively. From Co$^{3+}$ (3$d^6$) to Rh$^{3+}$ ions (4$d^6$), the cations at octahedral sites are in the NM state~\cite{42-1991-nishihara,27-Ohashi-1987}, leaving the AFM order constructed by magnetic Co($A$) ions (Fig.~\ref{XRD}(c)). Therefore, it is reasonable to consider that Co($A$) of Co$_{1.3}$Ir$_{1.7}$S$_4$ bears a similar AFM transition. In addition, the isothermal magnetization $M(H)$ displays a linear behavior both above and below $T_\mathrm{N}$ (see Fig. S3 in the SM~\cite{30-SM}).

The magnetic susceptibility of Co$_{1.3}$Ir$_{1.7}$S$_4$ at high temperatures was also measured (Fig.~\ref{Exp}(e)). We attempted to fit the data above 550 K with the extended Curie-Weiss formula $\chi = \chi_0 + C/(T - \theta_\mathrm{W})$, where $\chi_0$ is the sum of temperature-independent contributions; $C$ is the Curie constant; and $\theta_\mathrm{W}$ is the Weiss temperature. The tiny $\chi_0$, primarily contributed by core-electron diamagnetism and Van vleck paramagnetism, was omitted during fitting. The optimized fit yielded $C$ = 2.82 emu K mol$^{-1}$ and $\theta_\mathrm{W} = -469.18$ K, as indicated by the dash line in Fig.~\ref{Exp}(e). The effective magnetic moment $\mu_\mathrm{eff} =$ 4.75 $\mu_\mathrm{B}$ f.u.$^{-1}$, is significantly larger than that expected in the spin-only scenario (3.875 $\mu_\mathrm{B}$ f.u.$^{-1}$ for $S = 3/2$). Thus, there should be orbital contributions, as is reported in the analogous systems of CoAl$_2$O$_4$~\cite{24-roy-2013} and CoRh$_2$S$_4$~\cite{43-blasse-1965}. The frustration factor $f = \theta_\mathrm{W}/T_\mathrm{N} = 1.6$, suggesting a weak magnetic frustration in this system. The detailed magnetic configuration of Co$_{1.3}$Ir$_{1.7}$S$_4$ needs to be further clarified by neutron diffraction in the future. 

% C
To get more information about the magnetic transition, we measured the specific heat of Co$_{1.3}$Ir$_{1.7}$S$_4$ (Fig.~\ref{Exp}(c)). An obvious peak can be seen around 290 K, which corresponds to the AFM transition observed above. In the low-temperature regime, the specific heat was analyzed by the formula $C = \gamma T + \beta T^3$ shown in the inset of Fig.~\ref{Exp}(c), where $\gamma T$ represents the electron contribution and $\beta T^3$ represents phonon and AFM spin wave contribution. The fit gives $\gamma$ = 5.47 mJ K$^{-2}$ mol$^{-1}$ and $\beta$ = 0.406 mJ K$^{-4}$ mol$^{-1}$. Here, the non-zero $\gamma$ value is common in spinel insulators, which should be mainly attributed to lattice vacancies (approximately 3.8\% - 6\%)~\cite{44-schliesser-2015}.

% BAND-DOS
\begin{figure*}[t]
	\includegraphics[width=13cm]{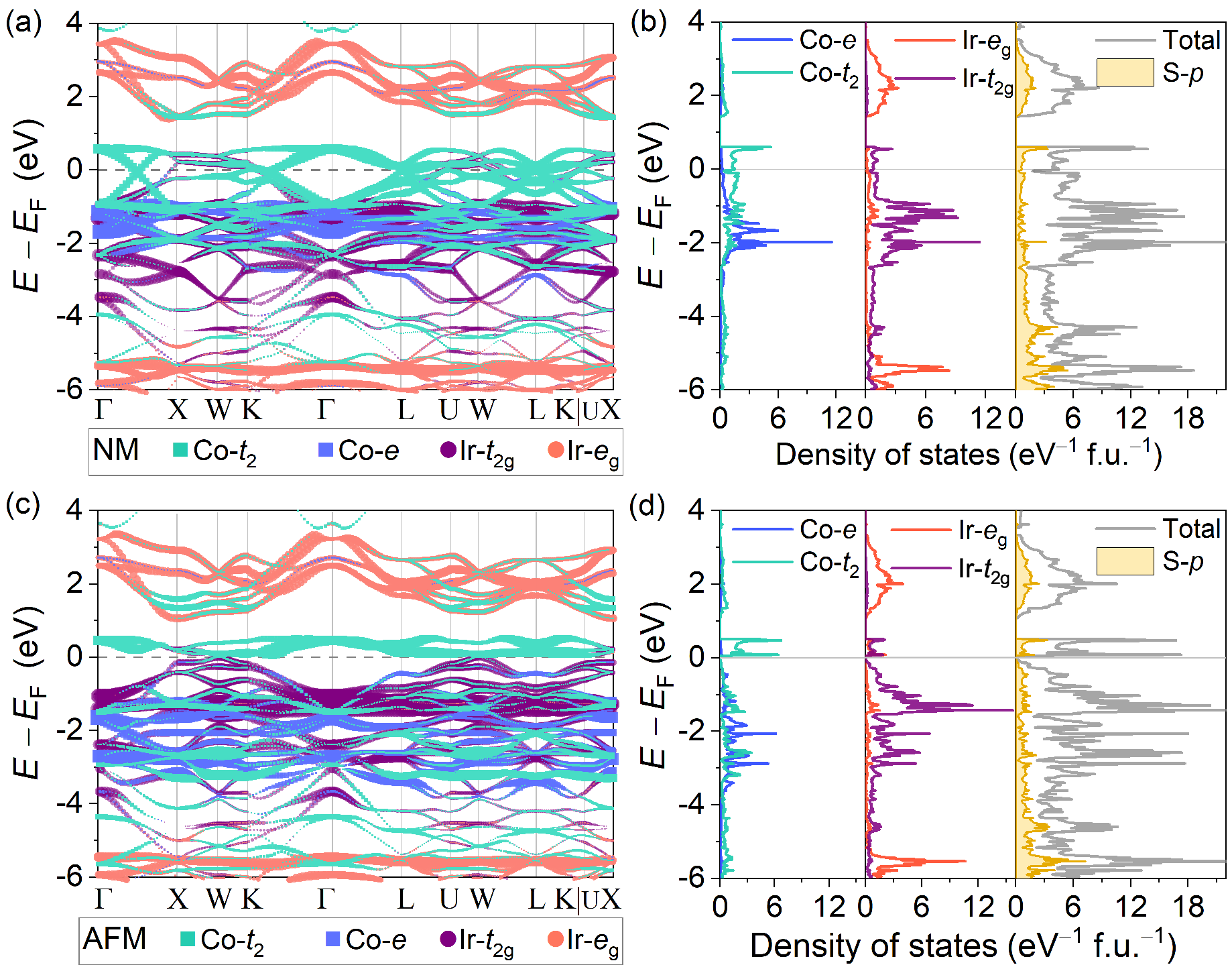}
	\caption{(Color online) (a) Electronic band structure projected onto different orbitals for nonmagnetic (NM) CoIr$_2$S$_4$. (b) Total and orbital-projected density of states for NM CoIr$_2$S$_4$. (c) Electronic band structure for antiferromagnetic (AFM) CoIr$_2$S$_4$. (d) Total and orbital-projected density of states for AFM CoIr$_2$S$_4$.} 
	\label{BAND-DOS}
\end{figure*}

In order to calculate the magnetic entropy change across the transition, we used a Debye-Einstein model to describe the phonon contribution of the total specific heat (Fig. S4 in the SM~\cite{30-SM}). After subtracting the background, we obtained the magnetic specific heat $C_\mathrm{m}$ and magnetic entropy $S_\mathrm{m}$, as shown in Fig.~\ref{Exp}(f). The total magnetic entropy of the AFM transition is $\sim$3.6 J K$^{-1}$ mol$^{-1}$, which is much lower than the expected value of $R\ln4$ for $S$ = 3/2. The reduced magnetic entropy is possibly due to short-range magnetic correlations above the magnetic transition, along with the strong Co-S hybridization (which will be shown later).

\subsection{\label{subsec:level3}Density-Functional Calculations.}

To further elucidate the magnetic and electronic properties, we performed first-principles calculations on the ideally nominal CoIr$_2$S$_4$. The electron correlation of cobalt element varies in chalcogenide systems~\cite{45-VAN-1991,46-kwon-2000,47-kim-2021}, thus we conducted study from the density functional theory (DFT) limit ($U$ = 0 eV) to the strongly correlated limit ($U$ = 4 eV) (Fig.~\ref{Phase_U}). We first calculated the total energies of different magnetic structures, where the AFM state consistently possesses the lowest energy at varied $U$ (Fig.~\ref{Phase_U}(a)). The magnetic moment of Co($A$) ions, $\mu_\mathrm{Co}$ increases monotonically with $U$. Note that in the supercell model of Co$_{10}$Ir$_{14}$S$_{32}$ (actually Co$_{1.25}$Ir$_{1.75}$S$_4$), the AFM ground state and $\mu_{\mathrm{Co}}$ remain, while no magnetic moment appears at the Co($B$) site, revealing that Co doping at $B$ site has tiny effect on this system. Meanwhile, CoIr$_2$S$_4$ undergoes a metal-to-insulator transition at $U =$ 1.5 eV, after which the band gap of the AFM insulator increases with larger $U$ (Fig.~\ref{Phase_U}(b), left axis). Comparing with experimental observations of the AFM insulating ground state with a band gap of 0.053(1) eV, we propose a reliable value of $U \sim$ 2 eV, as represented by the shaded region in Fig.~\ref{Phase_U}.

%Band-DOS
The band structure and density of states (DOS) for NM and AFM states of CoIr$_2$S$_4$ are presented in Fig.~\ref{BAND-DOS} (without specific clarification, the results refer to the calculations under $U$ = 2 eV). In the NM state, CoIr$_2$S$_4$ has a metallic band structure, where the states around Fermi level are dominantly contributed by Co-$t_2$, Ir-$t_\mathrm{2g}$ and S-$p$ orbitals. Meanwhile, the Co-$e$ orbitals have lower energies and form bands below Fermi level. The Ir-$e_\mathrm{g}$ orbitals, which strongly hybridize with the S-$p$ orbitals, split into two sets of bands, and locate well above and below Fermi level respectively. In the AFM state, an indirect gap emerges within the metallic bands (Fig. S5 in the SM~\cite{30-SM}). As we can see from Figs.~\ref{BAND-DOS}(c) and \ref{BAND-DOS}(d), when the band gap opens, the Co-$t_2$ components below Fermi level are significantly reduced. While the Co-$t_2$ components above $E_\mathrm{F}$ and the Ir-$t_\mathrm{2g}$ components below $E_\mathrm{F}$ show minor adjustments alongside the total DOS. Particularly, the number of unoccupied upper Co-$t_2$ bands in the AFM state is exactly 3, corresponding to the high spin 3$d^7$ configuration of the Co$^{2+}$ ions. However, a substantially reduced moment of 1.67 $\mu_\mathrm{B}$/Co (without the orbital moment) is observed, implying strong hybridization between Co-$t_2$ and S-$p$, which partially explains the small entropy change in Fig.~\ref{Exp}(f). 

%Schematic E
\begin{figure*}[t]
	\includegraphics[width=14cm]{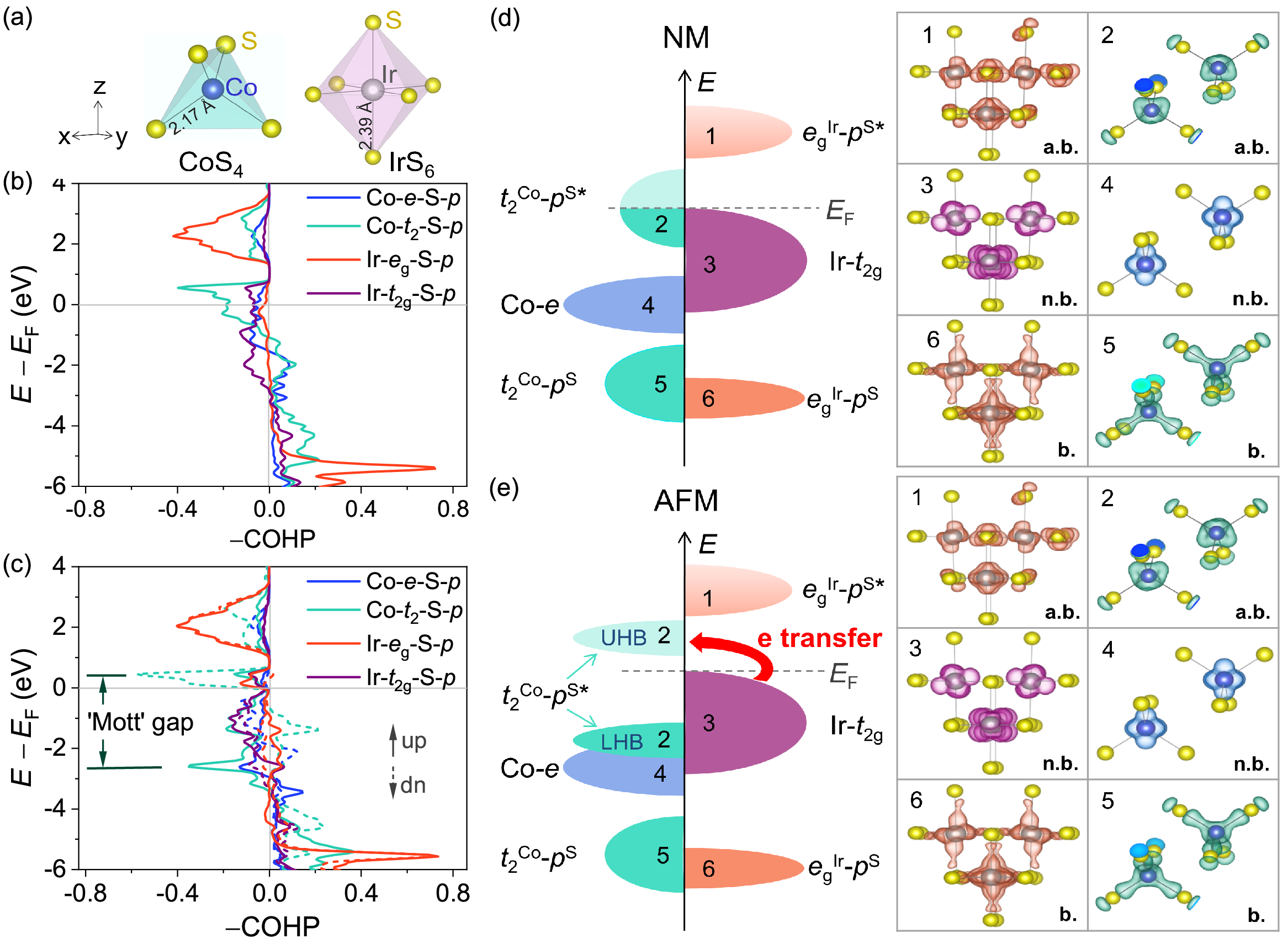}
	\caption{(Color online) (a) The CoS$_4$ tetrahedron, IrS$_6$ octahedron and corresponding local axes are shown. Crystal orbital Hamilton populations (COHP) analyses of the selected Co-S and Ir-S interactions for nonmagnetic (NM) CoIr$_2$S$_4$ (b) and antiferromagnetic (AFM) CoIr$_2$S$_4$ (c). The dash lines in (c) represent for spin down channel. (d) Schematic representation density of states (DOS) for NM CoIr$_2$S$_4$. Orange, turquoise, purple-red, and blue-violet colors represent the Ir-$e_\mathrm{g}$-S-$p$ (${e_\mathrm{g}}^\mathrm{Ir}$-$p^\mathrm{S}$) bonding and antibonding molecular orbitals (MOs), Co-$t_2$-S-$p$ (${t_2}^\mathrm{Co}$-$p^\mathrm{S}$) bonding and antibonding MOs, Ir-$t_\mathrm{2g}$ orbitals and Co-$e$ orbitals, respectively. Occupied states are shown in normal color and unoccupied states are remarked by lighter one. The Fermi energy is represented by a gray dashed line. The right panel is the corresponding charge density at $\Gamma$ point for these MOs. (e) Schematic representation DOS for AFM CoIr$_2$S$_4$, and the corresponding charge density at $\Gamma$ point for these orbitals are shown on the right. The ${t_2}^\mathrm{Co}$-$p^\mathrm{S}$ antibonding MOs split into lower Hubbard band (LHB) and upper Hubbard band (UHB). In (d) and (e), we annotate the antibonding, nonbonding and bonding states as a.b., n.b. and b., respectively.} 
	\label{NM-AFM}
\end{figure*} 

%COHP-PARCHG
The hybridization signals in DOS are further confirmed by the COHP analysis (Figs.~\ref{NM-AFM}(b) and \ref{NM-AFM}(c)). In both NM and AFM states, the Co-$t_2$-S-$p$ and Ir-$e_g$-S-$p$ orbital sets exhibit separated positive (or negative) peaks, demonstrating the formation of antibonding (or bonding) hybridized MOs. The Co-$e$-S-$p$ and Ir-$t_{2g}$-S-$p$ orbital sets, however, yield rather weak COHP values and show a nearly nonbonding nature. Spatially, the Co-$t_2$ and Ir-$e_\mathrm{g}$ orbitals have more significant overlaps with surrounding S-$p$ orbitals and thus stronger hybridization (Fig.~\ref{NM-AFM}(a)). Therefore, we can classify the energy states into six types: Ir-$e_\mathrm{g}$-S-$p$ bonding (${e_\mathrm{g}}^\mathrm{Ir}$-$p^\mathrm{S}$) or antibonding (${e_\mathrm{g}}^\mathrm{Ir}$-$p^\mathrm{S\ast}$) MOs, Co-$t_2$-S-$p$ bonding (${t_2}^\mathrm{Co}$-$p^\mathrm{S}$) or ${t_2}^\mathrm{Co}$-$p^\mathrm{S\ast}$ MOs, Ir-$t_\mathrm{2g}$ orbitals and Co-$e$ orbitals. Remarkably, from NM to AFM state, the spin polarization is dominated by the split of ${t_2}^\mathrm{Co}$-$p^\mathrm{S\ast}$ MOs. It is evident that such a splitting is accountable for the band gap in AFM CoIr$_2$S$_4$, highlighting the significant role of Mott physics within this system. 

Figures~\ref{NM-AFM}(d) and \ref{NM-AFM}(e) show the schematic view for the electronic structures of CoIr$_2$S$_4$ and corresponding charge distributions of typical $\Gamma$-point Bloch wave functions. The states within each band set are in good agreement with corresponding bonding, antibonding or nonbonding band characters. Meanwhile, these characters remain nearly unchanged from NM to AFM state, which indicate that the formation of ${t_2}^\mathrm{Co}$-$p^\mathrm{S\ast}$ MOs and Ir-$t_\mathrm{2g}$ orbitals could serve as a solid foundation for our theoretical framework. The ${t_2}^\mathrm{Co}$-$p^\mathrm{S\ast}$ MOs dominate the Fermi level in the NM state (Fig.~\ref{NM-AFM}(d), left panel), then split into lower and upper `Hubbard' bands (UHB) after the AFM order is switched on (Fig.~\ref{NM-AFM}(e), left panel). Since the band top of the Ir-$t_\mathrm{2g}$ is very close to the center of ${t_2}^\mathrm{Co}$-$p^\mathrm{S\ast}$ MO bands, the band splitting can easily pull the lower `Hubbard' band (LHB) below the band top of Ir-$t_\mathrm{2g}$. The energy gap between UHB and LHB, defined as the `Mott' gap, is always larger than the actual band gap of CoIr$_2$S$_4$ (Fig.~\ref{Phase_U}(b)). Thus, an ICCT band gap emerges between the upper ${t_2}^\mathrm{Co}$-$p^\mathrm{S}$* MO bands and the Ir-$t_\mathrm{2g}$ bands. 

%SE
\begin{figure*}[t]
	\includegraphics[width=14cm]{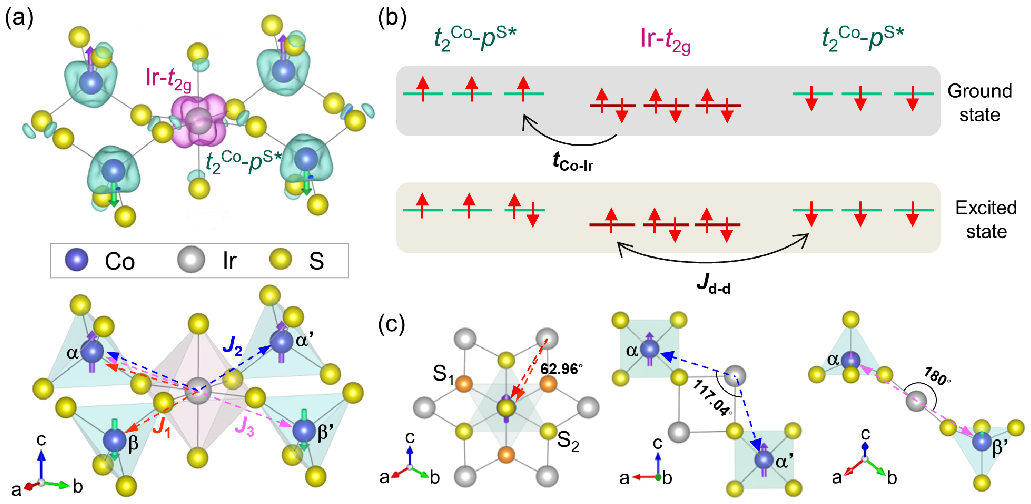}
	\caption{(Color online) (a) Illustration of long-range superexchange (SE) pathways in CoIr$_2$S$_4$. In the top half, the Co-$t_2$-S-$p$ antibonding molecular orbitals (${t_2}^\mathrm{Co}$-$p^\mathrm{S\ast}$) and Ir-$t_\mathrm{2g}$ orbitals are visually distinguished by fuchsia and turquoise colors, respectively. In the bottom half, the nearest neighbor (NN) exchange interactions $J_1$, next-nearest neighbor (NNN) exchange interactions $J_2$ and next-next-nearest neighbor (NNNN) exchange interactions $J_3$ are respectively indicated with red, blue and magenta dashed lines. Here, $\alpha$, $\beta$, $\alpha^{\prime}$ and $\beta^{\prime}$ represent the centers of CoS$_4$ complexes. $\beta$, $\alpha^{\prime}$ and $\beta^{\prime}$ are the NN, NNN and NNNN sites to $\alpha$, respectively. (b) Scheme of the long-range SE mechanism in CoIr$_2$S$_4$. (c) Six shortest exchange paths for $J_1$ (left), two shortest exchange paths for $J_2$ (middle) and one shortest exchange path for $J_3$ (right). In left panel, S belonging to different CoS$_4$ complexes are represented by varying colors.}
	\label{SE}
\end{figure*} 

\subsection{\label{subsec:level4}Long-Range SE Interactions.}
%SE
For CoIr$_2$S$_4$, the distance between the magnetic Co$^{2+}$ is 4.26 \AA, much larger than that in metal Co ($\sim$2.5 \AA)~\cite{48-hull-1921,49-kulesco-1968}, hence the direct exchange interaction is ignorable. Meanwhile, the ICCT band gap could be viewed as an analogy to the well-known charge-transfer gap in transition metal (TM) oxides~\cite{1-imada-1998}. This analogy allows for the construction of a theoretical model for long-range SE by substituting the TM-3$d$ and O-2$p$ orbitals with the ${t_2}^\mathrm{Co}$-$p^\mathrm{S\ast}$ MOs and the Ir-$t_\mathrm{2g}$ orbitals. Here in CoIr$_2$S$_4$, the fully occupied Ir-$t_\mathrm{2g}$ orbitals mediate the long-range SE interaction between two CoS$_4$ complexes, as depicted in Figs.~\ref{SE}(a) and \ref{SE}(b). The ligand S in the indirect exchange path~\cite{10-roth-1964,12-roth-1964}, is involved in the hybridization with Co($A$). However, in contrast to the speculation that magnetic exchange involves excitation to the empty $e_\mathrm{g}$ orbitals~\cite{10-roth-1964,28-mayer-1981}, our system demonstrates notably lower charge transfer energy from Ir-$t_\mathrm{2g}$ to ${t_2}^\mathrm{Co}$-$p^\mathrm{S\ast}$ states. Besides, the formation of hybridized MOs in CoIr$_2$S$_4$ weakens the local Coulomb interactions, as well as increases the hopping strength. Therefore one can expect significant enhancement of the long-range SE in CoIr$_2$S$_4$. 

Figure~\ref{SE}(a) illustrates the first-order SE interaction $J_1$ ($\alpha$-Ir-$\beta$), second-order SE interaction $J_2$ ($\alpha$-Ir-$\alpha^{\prime}$) and third-order SE interaction $J_3$ ($\alpha$-Ir-$\beta^{\prime}$). The nearest neighbor (NN) $J_1$ involves six equivalent Ir ions, while the next-nearest neighbor (NNN) $J_2$ involves two equivalent Ir ions, and the next-next-nearest neighbor (NNNN) interaction $J_3$ involves only one equivalent Ir ion (Fig.~\ref{SE}(c)). Although the angles of Co-Ir-Co increase sequentially from $J_1$ to $J_3$, these exchange interactions all appear to be AFM, which is different from the GKA rules~\cite{5-goodenough-1963}. In fact, due to the symmetry of the Ir tetrahedron, all Co-Ir-Co exchange paths give comparable and non-negligible hopping amplitudes. Therefore, the relative strengths of $J_1$, $J_2$, and $J_3$ mainly depend on their multiplicities. We attempted to use the Heisenberg spin Hamiltonian $H = -\sum_\mathrm{ij}J_\mathrm{ij}S_\mathrm{i}S_\mathrm{j}$ to describe the magnetic exchange interactions in this system. However, magnetic configurations other than ferromagnetic and AFM were unstable and converged to NM. Thus, only the combined AFM exchange coupling, $J = J_1 + 3J_3$, for CoIr$_2$S$_4$ is calculated to be $-19.4$ meV (details are shown in Fig. S6 of the SM~\cite{30-SM}). 

\begin{table}[b]
	\caption{Comparison of exchange-related parameters and magnetic properties of antiferromagnetic cobalt thiospinel Co$B_2$X$_4$ ($B$ = Al, Co, Rh, Ir; $X$ = O, S). Here, charge transfer energy $\Delta_{\mathrm{Co}-B}$ is derived from first principle calculations~\cite{50-Walsh-2007,51-smart-2019}, AFM exchange coefficient $J$ and N\'{e}el temperature $T_\mathrm{N}$ are obtained through experimental results~\cite{24-roy-2013,25-scheerlinck-1976,26-GeL.-2017,27-Ohashi-1987}.}
	\renewcommand\arraystretch{1.2}
	\begin{tabular}{ccccc}
		\hline\hline % Top horizontal line
		&  $\Delta_{\mathrm{Co}-B}$ (eV) & $J_\mathrm{eff}/t_{\mathrm{Co}-B}^4$ & $J$ (meV) & $T_\mathrm{N}$ (K)  \\
		\hline
		CoAl$_2$O$_4$ &  4.3   & 0.0664 & 0.92      & \textless 10 \\
		Co$_3$O$_4$   &  2.0   & 0.0853 & $\sim$0.6 & 40   \\
		CoRh$_2$O$_4$ &  1.8   & 0.0940 & 0.63      & 25  \\
		CoRh$_2$S$_4$ &  0.054 & 0.936  & -         & 418  \\
		CoIr$_2$S$_4$ &  0.045 & 0.946  & -         & 292  \\
		\hline\hline % Bottom horizontal line	
	\end{tabular}
	\label{Table2}
\end{table}

%SE
\begin{figure*}[t]
	\includegraphics[width=14cm]{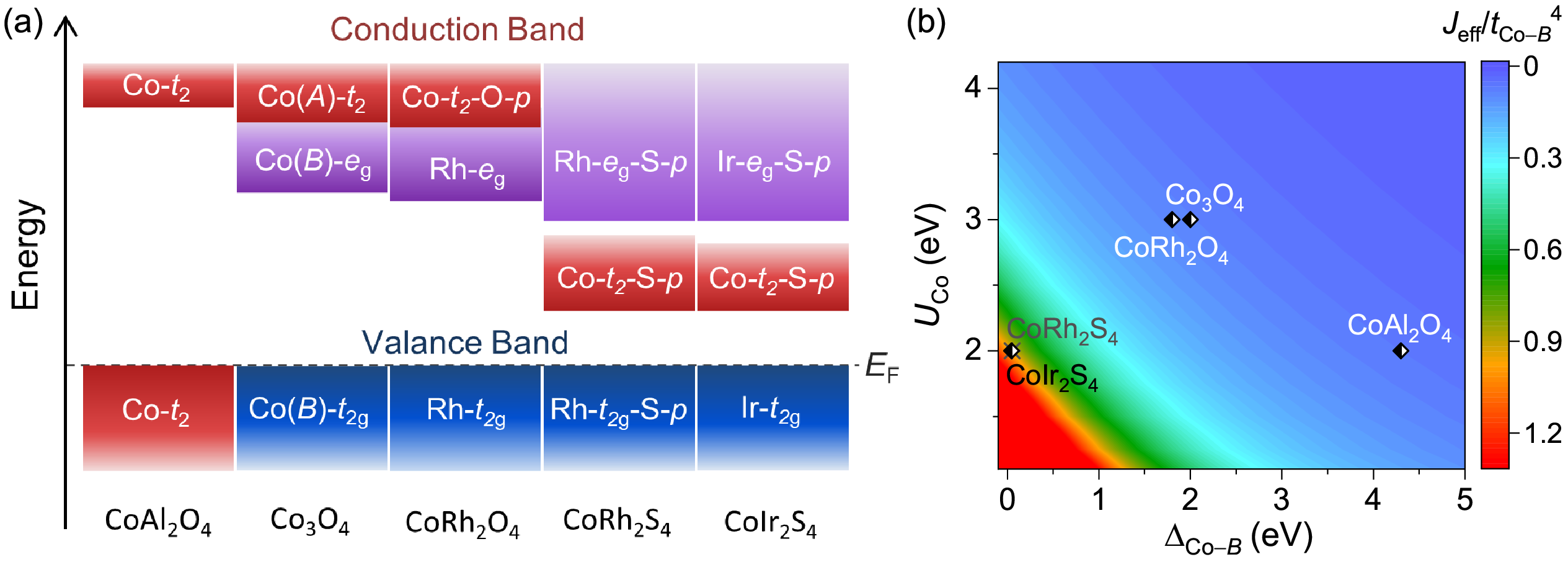}
	\caption{(Color online) (a) Schematic diagram of the band edges for antiferromagnetic cobalt thiospinel Co$B_2$X$_4$ ($B$ = Al, Co, Rh, Ir; $X$ = O, S). The electronic structures of CoAl$_2$O$_4$ and Co$_3$O$_4$ were from Refs~\cite{50-Walsh-2007,51-smart-2019}, and the electronic structures of CoRh$_2$O$_4$ and CoRh$_2$S$_4$ were calculated and illustrated in Fig. S7 of the Supplemental Material~\cite{30-SM}. The red, purple, mediumblue areas represents hybridized Co-$t_2$-$X$-$p$ (or Co-$t_2$), $B$-$e_\mathrm{g}$ and $B$-$t_\mathrm{2g}$ (or hybridized $B$-$t_\mathrm{2g}$-$X$-$p$) states, respectively. Note that the Fermi energies were set to zero. (b) The rescaled long-range superexchange coefficient $J_\mathrm{eff}/t_{\mathrm{Co}-B}^4$ as functions of the charge transfer energy $\Delta_{\mathrm{Co}-B}$ and the on-site Coulomb interaction $U_\mathrm{Co}$.} 
	\label{BAND}
\end{figure*} 

%Comparsion
The AFM exchange coefficient $\lvert J \rvert$ in CoIr$_2$S$_4$ demonstrates a notably larger magnitude compared to that of CoAl$_2$O$_4$~\cite{24-roy-2013}, Co$_3$O$_4$~\cite{25-scheerlinck-1976} and CoRh$_2$O$_4$~\cite{26-GeL.-2017}. To elucidate the differences, we summarized the valence band maximum (VBM) and conduction band minimum (CBM) of these Co$B_2X_4$ spinels, as depicted in Fig.~\ref{BAND}(a). Their electronic states at the band edges exhibit similarities, except for CoAl$_2$O$_4$, where Al-2$p$ occurs at a much lower energy level~\cite{50-Walsh-2007}. However, the charge transfer gap from the $B$-$t_\mathrm{2g}$ orbitals to the hybridized Co-$t_2$-$X$-$p$ (or Co-$t_2$) orbitals varies significantly among them. From the perturbation theory~\cite{1-imada-1998,52-eskes-1993}, the long-range SE interaction can be estimated: 
\[
J_\mathrm{eff} = \frac{4t^4_{\mathrm{Co}-B}}{(\Delta_{\mathrm{Co}-B} + U_\mathrm{Co})^2}(\frac{1}{U_\mathrm{Co}} + \frac{1}{\Delta_{\mathrm{Co}-B} + U_\mathrm{Co}}), 
\]
where $U_\mathrm{Co}$, $\Delta_{\mathrm{Co}-B}$ and $t_{\mathrm{Co}-B}$ is respectively the on-site Coulomb interaction, the charge transfer energy and the strength of Co-$B$ hopping. Thus, reducing $\Delta_{\mathrm{Co}-B}$ or $U_\mathrm{Co}$ will effectively strengthen the AFM coupling (Fig.~\ref{BAND}(b)). The large charge transfer energy of CoAl$_2$O$_4$, or the relatively strong on-site $U_\mathrm{Co}$ in Co$_3$O$_4$ and CoRh$_2$O$_4$, leads to a weak exchange coupling. While CoRh$_2$S$_4$ exhibits a similar $J_\mathrm{eff}/t_{\mathrm{Co}-B}^4$ value to CoIr$_2$S$_4$. From Table~\ref{Table2}, it can be observed that $J$ and $T_\mathrm{N}$ show positive correlation to the rescaled exchange coefficient $J_\mathrm{eff}/t_{\mathrm{Co}-B}^4$, confirming the applicability of the long-range SE model. Hence, the long-range SE interaction between magnetic Co($A$) mediated by $B$-$t_\mathrm{2g}$ orbitals is critical in determining the magnetic properties.

%Discussion
Unlike known SE~\cite{3-anderson-1959,6-zhang-2019,9-huang-2020}, the long-range SE interaction is distinctly mediated by cations. In addition to spinel, the involvement of cations in exchange pathways warrants careful consideration in materials with multiple cations, e.g. $AB$O$_3$ and $AA^{\prime}_3B_4$O$_{12}$ perovskite systems~\cite{53-azuma-2007,54-long-2009}. Especially, inter-cation (or intermetallic) charge transfer is observed within them under high pressure or temperature, accompanied by insulator-to-metal and antiferromagnetism-to-paramagnetism transitions~\cite{53-azuma-2007,54-long-2009,55-liu-2020,56-long-2012}. These results imply the highly probable long-range SE interactions facilitated by virtual inter-cation charge transfer at the ground state, consistent with the interplay of complex magnetic exchange in these materials~\cite{57-shimakawa-2015,58-huang-2014}. Overall, the insights gained from our study should be applicable to other materials with multiple cations that potentially bear long-range SE mechanism. 

\section{\label{sec:level4}Concluding Remarks}

In summary, we have synthesized and systematically investigated the physical properties and electronic structure of the new thiospinel Co$_{1+x}$Ir$_{2-x}$S$_4$ ($x$ = 0.3). Electrical transport measurement shows that it is an insulator, with a band gap of 0.053(1) eV. Magnetic susceptibility and specific heat measurements demonstrate that Co$_{1+x}$Ir$_{2-x}$S$_4$ undergoes an AFM transition at $T_\mathrm{N} \sim$ 292 K. The ordered moment is predicted to be 1.67 $\mu_\mathrm{B}$/Co($A$) theoretically. Our DFT calculations, with inclusion of Hubbard $U$ = 2 eV for Co-3$d$, yield consistent results with experimental observations in nominal CoIr$_2$S$_4$. An ICCT band gap between ${t_2}^\mathrm{Co}$-$p^\mathrm{S\ast}$ and Ir-$t_\mathrm{2g}$ states is formed in the AFM ground state. Utilizing the analysis of MO, we proposed a robust long-range SE interaction between NN Co($A$) mediated by Ir-$t_\mathrm{2g}$ orbitals. The large energy scale of the AFM exchange interaction ($-19.4$ meV) accounts for the high $T_\mathrm{N}$ observed in Co$_{1+x}$Ir$_{2-x}$S$_4$. Our results lay the groundwork for exploring the long-range SE interaction in ICCT insulators.

%%%%%%%%%%%%%%%%%%%%%%%%%%%%%%%%%%%%%%%%%%%%%%%%%%%%%%%%%%%%%%%%%%%%%
%% The "Acknowledgement" section can be given in all manuscript
%% classes.  This should be given within the "acknowledgement"
%% environment, which will make the correct section or running title.
%%%%%%%%%%%%%%%%%%%%%%%%%%%%%%%%%%%%%%%%%%%%%%%%%%%%%%%%%%%%%%%%%%%%%
\begin{acknowledgements}
	
This work was supported by the National Key Research and Development Program of China (2023YFA1406101, 2022YFA1403202), and the National Natural Science Foundation of China (12050003), and the Key Research and Development Program of Zhejiang Province, China (2021C01002). 

Liang-Wen Ji and Si-Qi Wu contributed equally to this work.	
\end{acknowledgements}

%%%%%%%%%%%%%%%%%%%%%%%%%%%%%%%%%%%%%%%%%%%%%%%%%%%%%%%%%%%%%%%%%%%%%
%% The appropriate \bibliography command should be placed here.
%% Notice that the class file automatically sets \bibliographystyle
%% and also names the section correctly.
%%%%%%%%%%%%%%%%%%%%%%%%%%%%%%%%%%%%%%%%%%%%%%%%%%%%%%%%%%%%%%%%%%%%%

\section*{References}

\normalem

\bibliographystyle{apsrev4-2}
%\bibliography{CoIr2S4}

%apsrev4-2.bst 2019-01-14 (MD) hand-edited version of apsrev4-1.bst
%Control: key (0)
%Control: author (72) initials jnrlst
%Control: editor formatted (1) identically to author
%Control: production of article title (-1) disabled
%Control: page (0) single
%Control: year (1) truncated
%Control: production of eprint (0) enabled
%

\end{document}